# Real-Time Dynamic Data Driven Deformable Registration for Image-Guided Neurosurgery: Computational Aspects

Nikos Chrisochoides[1], Andrey Fedorov[1,3], Yixun Liu[1], Andriy Kot[1], Panos Foteinos[1], Fotis Drakopoulos[1], Christos Tsolakis[1], Emmanuel Billias[1], Olivier Clatz[2], Nicholas Ayache[2], Alex Golby[3,4], Peter Black[4] and Ron Kikinis[3]

[1]Center for Real-Time Computing, Computer Science Department, Old Dominion University, Norfolk, VA
[2] Inria, French Research Institute for Digital Science, Sophia Antipolis, France
[3]Neuroimaging Analysis Center, Department of Radiology, Harvard Medical School, Boston, MA
[4]Image-guided Neurosurgery, Department of Neurosurgery, Harvard Medical School, Boston, MA

**Abstract** Current neurosurgical procedures utilize medical images of various modalities to enable the precise location of tumors and critical brain structures to plan accurate brain tumor resection. The difficulty of using preoperative images during the surgery is caused by the intra-operative deformation of the brain tissue (brain shift), which introduces discrepancies concerning the preoperative configuration. Intra-operative imaging allows tracking such deformations but cannot fully substitute for the quality of the pre-operative data. Dynamic Data Driven Deformable Non-Rigid Registration (D$^4$NRR) is a complex and time-consuming image processing operation that allows the dynamic adjustment of the pre-operative image data to account for intra-operative brain shift during the surgery. This paper summarizes the computational aspects of a specific adaptive numerical approximation method and its variations for registering brain MRIs. It outlines its evolution over the last 15 years and identifies new directions for the computational aspects of the technique.

## 1. Introduction

Cancer continues to be a significant cause of death in the USA and worldwide. The number of Americans living with brain tumors exceeds 700,000, surpassing the number of COVID-19 deaths in mid-summer 2021. Neurosurgical resection is a standard and effective treatment for brain tumor patients. Removing as much of the tumor as possible is imperative to ensure the best results while preserving healthy brain structures. This approach can extend the progression time while reducing symptoms and seizures.

One of the main challenges in neurosurgery is identifying critical areas of the brain responsible for essential functions, such as the motor cortex.

These areas are unique to each patient and cannot be located with the naked eye. However, medical imaging has proven to be an asset in overcoming this hurdle. Over the past two decades, advancements in image-guided therapy (Grimson et al., 1999) have allowed surgeons to utilize preoperative imaging (Orringer et al., 2012) for neuronavigation. With visualization (Kikinis et al., 2014) and quantitative analysis software systems (Fedorov et al., 2012), surgeons can safely remove tumors, such as gliomas, from sensitive brain areas. These advancements have significantly improved neurosurgery's safety and success rates.

Before surgery, a combination of anatomical Magnetic Resonance Imaging (MRI) and functional MRI (fMRI) can pinpoint crucial brain areas that affect functions such as vision, speech and language, or motor control. Moreover, Diffusion Tensor Imaging (DTI) can map out white matter tracts that connect to these essential regions and are located near or through the tumor. These imaging techniques are essential in ensuring the precision of the tumor removal procedure.

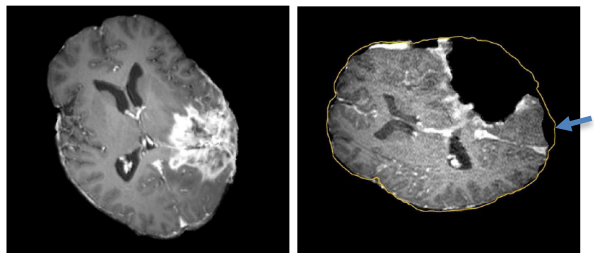

Figure 1. Discrepancies between preoperative and intraoperative MR Imaging before and during neurosurgery. (Left): preoperative MRI; (Right): intraoperative MRI acquired after a part of the tumor is removed shows the tumor cavity and brain shift (pointed by the blue arrow). The yellow outline indicates the preoperative brain outline after rigid registration. The large dark cavity is due to tumor resection.

During surgery, the opening of the skull and dura causes changes in pressure inside the Intra-Cranial Cavity. Because of this and other factors, such as drainage of cerebrospinal fluid and the effect of gravity, the brain changes its shape, introducing discrepancies in relation to the pre-operative configuration. The adoption of intraoperative MRI (iMRI) has provided a means for monitoring brain deformation (or brain shift) during surgery (Golby, 2015). The number of hospitals offering iMRI has grown over the past decade from a handful of research centers to hundreds of clinical sites worldwide (Black, 2016). iMRI can be used to observe the deformed brain during surgery. While acquiring fMRI and DTI is intraoperatively impractical, the preoperative MRI image can be registered to an iMRI using non-rigid registration. The resultant registration can then be applied to the preoperative fMRI, DTI, and surgical plan, providing more accurate, updated guidance to the neurosurgeon (Archip et al., 2007).

Image registration, in general, is concerned with the spatial alignment of corresponding features in two or more images. During image registration, a spatial transformation is applied to one image (called floating) so that it is brought into alignment with the fixed or target image, which is used as a reference position of the object (patient's brain). In this method, two types of image registration are used. First, the floating image, which corresponds to the pre-operative MRI, is aligned with the patient's position using translations and rotations (i.e., global transformations). Second, *non-rigid* registration uses spatially varying (i.e., local) transformation to account for brain shift, which varies in different brain locations. Image registration algorithms generally optimize specific similarity criteria between the fixed and floating image under varying spatial transformation parameters. The complexity of this optimization depends on the number of parameters that describe the transformation. Non-rigid registration usually requires significant computing resources and time; it is an open research area in medical image processing and a potential use case for quantum advantage even in the Noisy Intermediate-Scale Quantum (NISQ) era, Figure 2 depicts the fusion of structural and functional MRI near a brain tumor, a tessellation, and the computed deformation field superimposed over iMRI (top right).

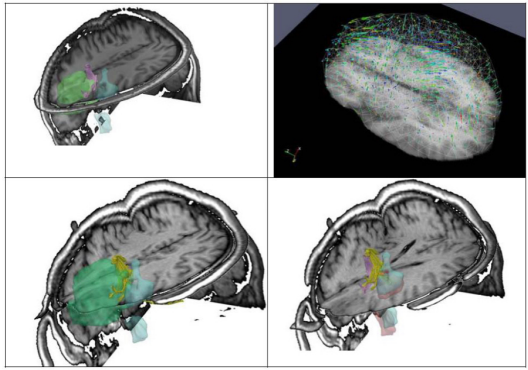

Figure 2. Brain tumor (green) and fMRI fused with iMRI using a deformation field (top right).

Figure 3 depicts a flowchart with all steps and software modules for pre- and intra-operative image processing for image-guided neurosurgery at Brigham and Women's Hospital (BWH) in Boston, MA. The intra-operative images were 0.5T iMRI (J. Schenck, F. Jolesz, P. Roemer, H. Cline, W. Lorensen, R. Kikinis, et al., 1995) and acquired during surgery at BWH. The patient-specific Finite Element (FE) model, selection of registration points, and non-rigid registration took place remotely at the Center for Real-Time Computing (CRTC) in Virginia using midsize High-Performance Computing (HPC) clusters (Chrisochoides et al., 2006).

This paper presents an overview of the physics-based non-rigid registration method (Clatz et al., 2005) and briefly describes its extensions to improve its accuracy for large brain tumors. Namely, it presents the original method developed by Clatz et al. at INRIA, France, and its implementation within Insight Segmentation and Registration Toolkit (Liu and Kot et al., 2014). It overviews its extensions by removing additional outliers due to tissue resection (Liu et al., 2014) and (Drakopoulos et al., 2015) at the CRTC in collaboration with the Surgical Planning Lab (SPL) at BWH. It is outside the scope of this paper to review past work by others; however, in (Sotiras et al., 2013), a complete survey and taxonomy of NRR methods are presented. The aim is to identify opportunities for leveraging quantum computing and sensing. Namely, we look for a hybrid classical and quantum computing paradigm to improve computational intensive modules identified in the rest of the paper and the potential use of quantum sensing to eliminate the high cost of encoding classical DICOM images to quantum computers.

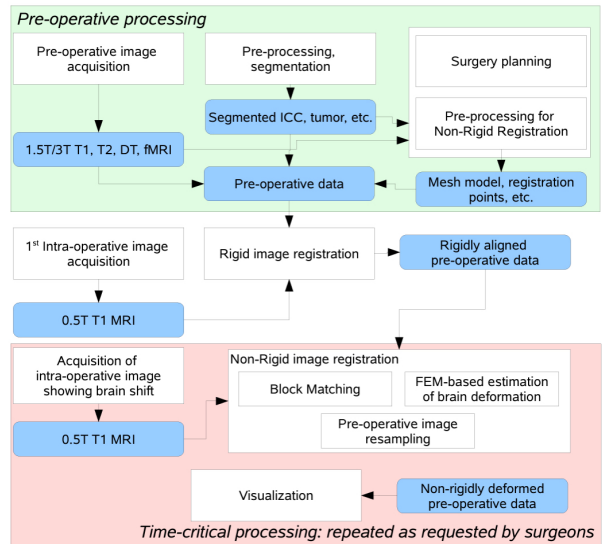

Figure 3. Flow diagram of the NRR process used in BWH during a clinical study, as depicted in Figure 3 (Archip et al., 2007).

## 2. Physics-Based Non-Rigid Registration

The specific NRR method was initially developed in INRIA, France, by Clatz and Ayache et al. (Clatz et al., 2005) and is implemented as open-source software by the CRTC group in Virginia, USA (Liu and Kot et al., 2014). It is designed for registering high-resolution pre-operative data with iMRI—the NRR process takes place in two phases: preoperative and intra-operative. Intra-operative processing (i.e., the dynamic data-driven stage) starts with the acquisition of the first iMRI, i.e., the *time-critical DDDAS* part, and the intra-operative computation is initiated when a scan shows the shift of the brain becomes available. The basic idea of the registration method is to estimate the *sparse deformation field* that matches "similar" locations in the pre-operative and iMRI and then use a biomechanical model for brain deformation to discard unrealistic displacements so that it can derive a *dense deformation field* that defines a transformation for each point in the image space.

Sparse displacement vectors are obtained at the selected points in the image, where the intensity variability in the surrounding region exceeds a certain threshold. Such *registration (or feature)* points can be identified before the time-critical part of the computation in the floating (pre-operative) image. Once the reference (intraoperative) scan is available, the deformation vector is estimated at each of the selected points utilizing block matching (Clatz et al., 2005), where fixed-size rectangular regions (blocks) centered at the registration points are identified in the floating image. Given such a block, the method selects a search region (window) in the reference image. At the registration point, the vector of the sparse deformation field is defined by the block's displacement, which produces the most significant similarity between the image intensities in the block and the overlapping section of the window. The normalized cross-correlation similarity metric is used. It is worth noticing the high computational complexity of the block-matching procedure. Considering the sizes of three-dimensional block and window are defined in pixels as $B = \{Bx, By, Bz\}$ and $W = \{Wx, Wy, Wz\}$, the bound on the number of operations is $O(BxByBz \times WxWyWz)$ for a single registration point.

This problem can be formulated as energy minimization (Clatz et al., 2005). One seeks the balance between the external forces, proportional to the sparse displacements, and the internal forces of the mesh resisting deformation:

$$[K + H^T SH]U = H^T SD \tag{1}$$

*where K* is the mesh stiffness matrix, *H* is the linear interpolation matrix form the matches to the displacements at mesh vertices, *S* is the block matching stiffness matrix (matches with higher confidence are assigned higher weights), *D* is the vector for the block displacements, and *U* is the unknown displacement vector for mesh vertices. The stiffness matrix, *K*, is calculated based on the assumed physical properties of the brain tissue elastic modulus *E* and Poisson ratio ν (see Table 5). This formulation can tolerate some outliers but suffers from a systematic error concerning the correctly estimated displacements. Alternately, one can use approximation to compute the locations of vertices, which would minimize the error concerning the block matches:

$$\arg\min_U (HU - D)^T S(HU - D) \tag{2}$$

However, this formulation would also minimize displacement error regarding outlier measurements, which one would like to eliminate from the set of block displacements. A robust iterative approach combines approximation and interpolation. Gradual convergence to the interpolation solution is achieved using the external force *F* added to the formulation (1) to slowly relax the internal mesh stress:

$$[K + H^T SH]U = H^T SD + F \qquad (3)$$

Rejection of the outlier matches is done iteratively, with a user-defined total percentage of matches to be discarded, $f_R$, and the number of rejection iterations, $n_R$, as follows:

```
1:  INPUT: n_R, f_R
2:  for i =0 to n_R do
3:      F_i ⇐ KU_i
4:      U_{i+1} ⇐ [K + H^T SH]^{-1} [H^T SD + F_i]
5:      for all blocks k do
6:          compute error function ξ_k
7:      end for
8:      reject f_R/ n_R blocks with highest error function ξ
9:      re-compute S, H, D
10:     end for
11: repeat
12:     F_i ⇐ KU_i
13:     U_{i+1} ⇐ [K + H^T SH]^{-1} [H^T SD + F_i]
14: until convergence
```

The force, *F*, is computed at each iteration to balance the internal force of the mesh, $KU_i$. The error, $\xi_k$, measures the difference between the block displacement approximated from the current deformed mesh and the matching target for the *k*th block. The user-defined percentage of the displacements with the highest $\xi_k$ values is rejected. This method converges to the formulation in (2) and, at the same time, is tolerant to most of the outliers due to faulty matching.

The Physics-Based Non-Rigid Registration (PBNRR) method is implemented (Liu and Kot et al., 2014) as part of the Insight Segmentation and Registration Toolkit (ITK) 4.5 refactoring effort sponsored by the National Library of Medicine. ITK is a multi-platform, open-source image analysis library serving many researchers and engineers worldwide. ITK collects cutting-edge image analysis algorithms, providing a platform for advanced product development.

The PBNRR includes three main modules components, all of which have been implemented in ITK:
- Feature Point Detection: Identify small image blocks with rich structural information in the pre-operative MRI.
- Block Matching: Calculate displacement for each image block to generate a sparse deformation field.

- Robust Finite Element (FE) Solver: Estimate entire brain deformation based on the sparse deformation field estimated above.

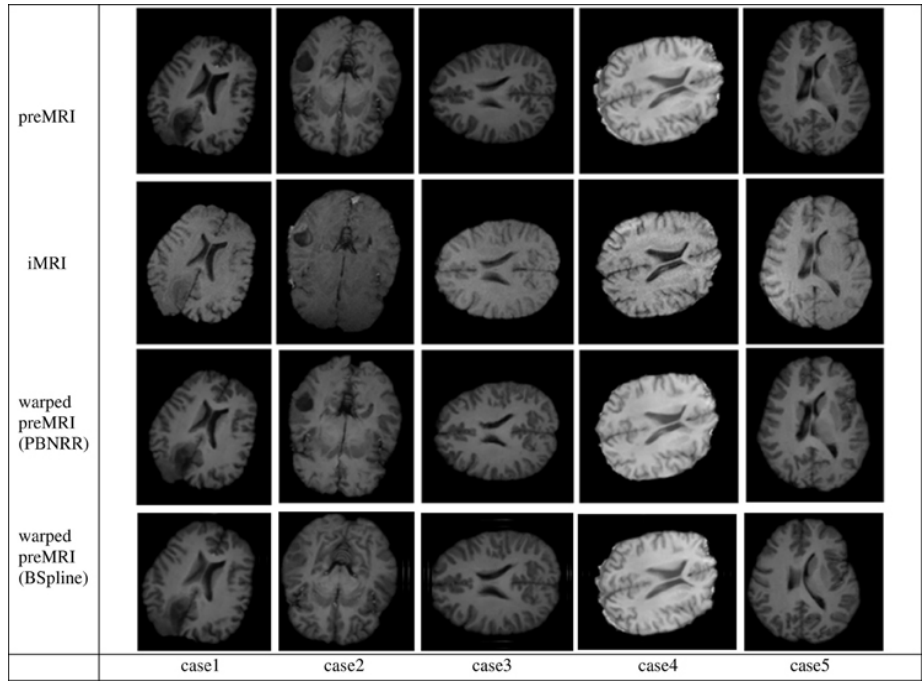

Figure 4. Non-Rigid Registration of pre-operative (top row) with iMRI (2nd row) using PBNRR and B-Splines for four cases included in the (NCIGT, 2021) dataset.

Studies conducted by Clatz et al. in 2005 and independently by Liu and Kot et al. in 2014, Liu et al. in 2014, and Drakopoulos et al. in 2021 have demonstrated that the biomechanical model presented by Clatz et al. and its ITK implementation generally provide reasonable estimates of brain shift in the context of sparse and "noisy" intra-operative data. This is particularly true just after the dura is opened and before any tissue is removed when compared to the standard registration method used in most Image Guided Navigation Systems (IGNS), known as rigid registration. In addition, PBNRR usually performs better than the best interpolation methods, e.g., the B-Spline-based registration method implemented within Slicer3D. The reason is that PBNRR can: (i) *capture local brain shift* better because of its natural piecewise approximation as opposed to global affine transformations utilized in rigid registration and (ii) remove some of the outliers in contrast to interpolation methods, which by design use all data independently of the quality of information (i.e., being low or high confidence). Both B-Splines and PBNRR are currently available within Slicer3D.

In (Liu et al., 2010), we confirmed that the PBNRR method is ready for wide application in the OR without requiring distributed or cluster computing resources, as was the case in (Chrisochoides et al. l, 2006). However, no computational power can improve the method's accuracy for the complete resection of large brain tumor cases leading to significant brain shifts. This creates a large cavity of elements in the tessellated brain image model (as depicted in Figure 4), which compromises the accuracy of the biomechanical model defined on pre-operative MRI, given that this cavity is not properly accounted for.

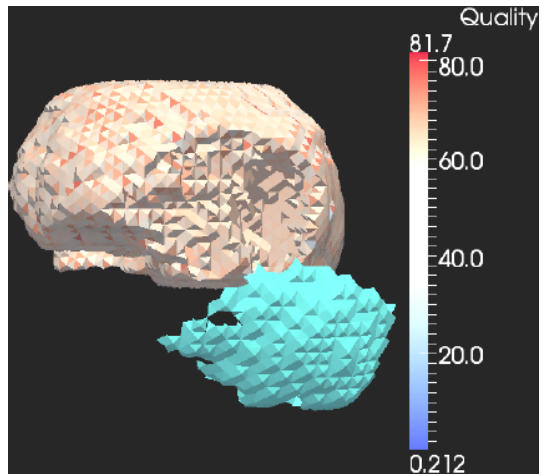

Figure 5. Tesselatted brain image and cavity of brain tumor elements (green).

## 3. Real-Time D$^4$NRR

This section summarizes the extensions of the PBNRR by (i) removing additional outliers due to tissue resection using either algebraic means: (1) a Nested Expectation and Maximization method (referred to thought-out as NEMNRR) (Liu et al., 2014), or (2) geometric by relying on real-time image-to-mesh tessellation method (Foteinos et al., 2014) and (ii) considering a heterogeneous biomechanical model as opposed to homogenous used in (Clatz et al., 2005). Follow-up efforts by this group will integrate these new capabilities within the next major upgrade of ITK and Slicer 3D.

### 3.1 Nested Expectation Maximization Method

The NEMNRR method formulates the registration as a three-variable (point correspondence, deformation field, and resection region) functional minimization problem, in which point correspondence is represented by a fuzzy assign matrix; the Deformation field is represented by a piece-wise linear function regularized by the strain energy as in PBNRR (Clatz et al., 2005), but this time extends the model from a single homogenous tissue to a heterogeneous multi-tissue (see Figure 6) based biomechanical model. A Nested Expectation and Maximization framework is developed to resolve these three variables simultaneously (Liu et al., 2014).

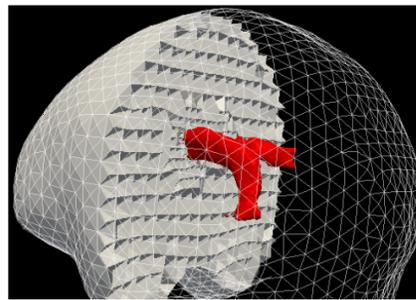

Figure 6. Multi-tissue mesh from segmented brain MRI.

The NEMNRR method extends the cost function used in (Clatz et al., 2005) to:

$$J(U, C, M_{Rem}) = \sum_{e_i \in M \setminus M_{Rem}} U^T K_{e_i} U +$$

$$\lambda_1 \sum_{s_i \in M \setminus M_{Rem}} (HU - D(C))^T W (HU - D(C)) \qquad (4)$$

$$+ \lambda_2 \sum_{e_i \in M_{Rem}} V_{e_i}$$

where the continuous domain $\Omega$ (brain image) is discretized as a multi-tissue mesh $M$ using the method presented in (Liu et al., 2010) on a multi-label image segmented from the pre-operative MRI. $M_{Rem}$ is the removed mesh approximating the resection region $\Omega'$. $K_{e_i}$ is the element stiffness matrix of element $e_i$. Each element is associated with a tissue label, which determines the elastic parameters to build the element stiffness matrix. The first term of equation (4) approximates the strain energy as in (Clatz et al., 2005), and the third term approximates the volume of the resection region, in which $V_{e_i}$ is the volume of element $e_i$. In the second term, the entries of the vector $D$ are defined as

$$d_i(c_{ij}) = s_i - \sum_{t_j \in \Omega_R} c_{ij} t_j, \forall s_i \in M \setminus M_{Rem}.$$

Considering the registration problem in the Expectation and Maximization (EM) context (Dempster et al., 1977), cost function (4), from the probability (Bayesian) point of view, defines the likelihood function, in which the unknown (model parameter) is the displacement vector $U$, and the missing data are the correspondence $C$ and the resection region $M_{Rem}$. Assuming $M_{Rem}$ is known, the more accurate the estimate of $C$, the more accurate the estimate of $U$, and vice versa. EM algorithm is very efficient for this kind of circular dependence problem, so one employs EM to solve $U$ and $C$ under a specified $M_{Rem}$. To resolve $M_{Rem}$, one can treat $U$ and $C$ as an unknown pair $\langle U, C \rangle$. The more accurate the estimate of $M_{Rem}$, the more accurate the estimate of $\langle U, C \rangle$, leading to a Nested EM framework as shown in Figure 7, in which the inner EM serves to resolve $\langle U, C \rangle$ with $M_{Rem}$ fixed, and the outer EM serves to resolve $M_{Rem}$. $M_{Rem}$ is approximated by a collection of tetrahedra located in a region of the model, which corresponds to the resection region in the intraoperative MRI. $M_{Rem}$ is initialized to $\emptyset$ and updated at each iteration of the outer EM. The outer EM stops if all the tetrahedra contained in the resection region are collected.

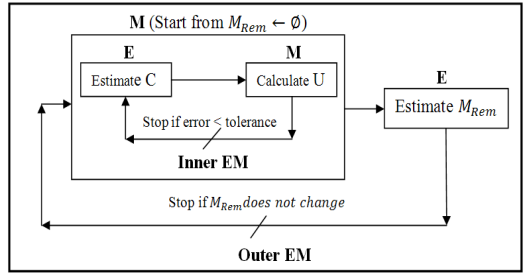

Figure 7. Nested Expectation and Maximization framework.

The resection region is difficult to identify in the intra-operative MRI, so a simple threshold segmentation method is used. We cannot determine if a tetrahedron is an outlier based solely on its position. It might be in the background image (BGI) instead of the resection region. To ensure the element outlier rejection algorithm is robust, we use the fact that the resection region is made up of tetrahedra that not only fall in the BGI of intra-operative MRI but also connect and form a maximal connected submesh. The outliers are collected iteratively, with additional outliers added into $M_{Rem}$ if they fall in the BGI and connect with the maximal simply connected submesh identified in the previous iteration. We demonstrate the NEMNRR strategy in Figure 8, with the inner EM iterating horizontally and the outer EM iterating vertically.

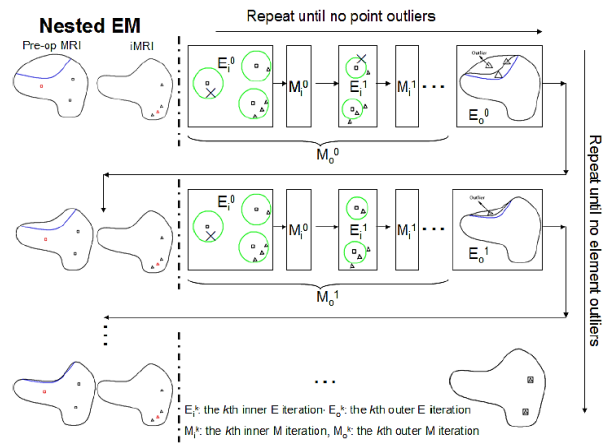

Figure 8. Illustration of Nested Expectation and Maximization strategy. Row: inner EM, Column: outer EM.

NEMNRR addressed a fundamental challenge in PBNRR: *"It does not require the point correspondence to be known in advance and allows the input data to be incomplete, thus producing a more general point-based NRR method."* This implies that pre-operative landmarks near the tumor fail to correspond to iMRI landmarks and thus are treated as outliers, and the corresponding elements of the mesh are marked and removed. The crux of the idea is to use the NEM method to resolve the deformation field with missing correspondence, specifically in the resection region. *This has many implications; one is to compute the registration error more accurately than HD, which will be explored in the future using quantum computing.* Like the PBNRR, the NEMNRR uses the strain energy of the biomechanical model to regularize the solution. As a result, it is improving the registration error over the PBNRR by 2.9mm on average out of 25 cases presented in (Liu et al., 2014).

The NEMNRR addresses brain deformation caused by tumor removal, but it is computationally expensive and currently is very difficult to meet clinical time constraints, even with the use of the best available accelerators. On the other hand, it does not require advanced knowledge of point correspondence and can handle incomplete input data, resulting in a more comprehensive point-based NRR. This method utilizes the biomechanical model's strain energy to regulate the solution and utilizes a multi-tissue heterogeneous model to enhance the simulation's accuracy, while the PBNRR relies on a simpler homogenous model.

*3.2 Real-Time Incremental Adaptive Deformable Registration*

Despite the qualitative improvements made by NEMNRR, the registration error for cases with extensive tumor resections was still beyond the threshold of 1mm to be acceptable for routine use in the operating room. The next attempt in (Drakopoulos et al., 2014) used geometric means to remove outliers due to extensive tumor resections (as in NEMNRR), while in the meantime, the CRTC group built real-time multi-tissue I2M conversion technologies (Foteinos, 2014) that permitted the re-meshing of the brain multi-labeled image in less than a second using multi-core hardware and thus able to address the time critical aspects of adaptivity in D$^4$NRR

The Adaptive Non-Rigid Registration (ANRR) method gradually adjusts the mesh for the FEM model to an incrementally warped segmented iMRI as opposed to NEMNRR that iteratively rejects feature and element outliers derived from a single (original) segmented iMRI. The idea of the geometric approach is to remove slivers and potentially negative volume elements resulting from large deformation fields (sometimes larger than the size of the elements) computed by block matching. This is achieved through an incremental approximation to reach the end goal. The ANRR method improves the accuracy of the model by improving the accuracy of the basic numerical calculations involved at the cost of increasing (potentially) the overhead for the mesh generation step and substantially increasing the computational cost of the linear solver several times. However, even with a single HPC node (DELL workstation with 12 Intel Xeon X5690@3.47 GHz CPU cores and 96 GB of RAM), the *ANRR execution time on average is less than two minutes (Drakopoulos et al., 2017), which is within the time constraints of the procedure in the operating room.*

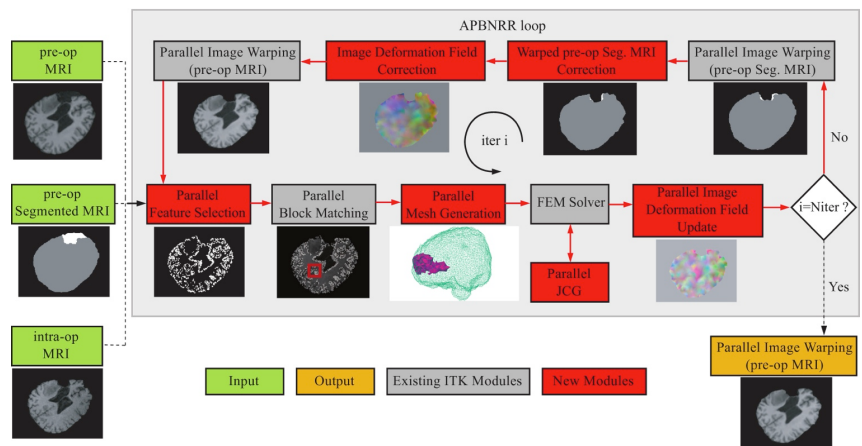

Figure 9. The ANRR software architecture. The red arrows show the execution order of the different modules. The loop breaks when the desired number of iterations has been reached. The output image (orange box) is the warped pre-op MRI when i = $N_{iter}$.

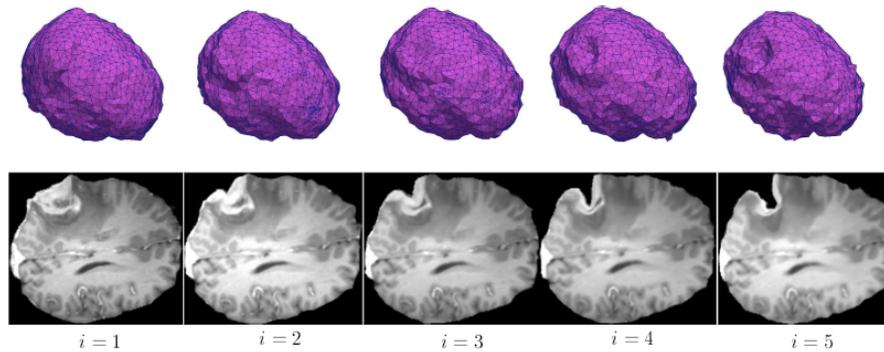

Figure 10. Non-rigid registration using five iterations of ANRR reflects the changes in brain morphology caused by tumor resection. Each column depicts a brain mesh model (top) and an axial slice of the preoperative image after the non-rigid registration (bottom) at iteration i.

The evaluation from two retrospective studies using initially 14 and then 30 cases indicated that the geometric scheme is reducing substantially the alignment error compared to NEMNRR, PBNRR, B-Spline, and Rigid registration (Drakopoulos et al., 2014 and 2021).

Table 1. Qualitative data of the intermediate adaptive mesh refinement for ANRR.

| Adaptive Iteration | #Tets | $\alpha_{min}$ | | $\alpha_{max}$ | |
|---|---|---|---|---|---|
| | | Before | After | Before | After |
| 1 | 14278 | - | 5.00° | - | 169.68° |
| 2 | 13482 | 0.34° | 5.24° | 179.28° | 169.92° |
| 3 | 13497 | 0.14° | 4.91° | 179.79° | 169.71° |
| 4 | 12957 | 0.10° | 5.01° | 179.80° | 171.32° |

The error in these studies and this paper (unless it is stated otherwise) is measured in terms of Hausdorff Distance (HD) as the measurement of the degree of mismatch between two pointsets with the equation,

$$H(A,B) = \max[h(A,B), h(B,A)] \qquad (5)$$

where $h(A,B)$ and $h(B,A)$ directed HD defined by $h(A,B) = \max_{a \in A} \min_{b \in B} \|a-b\|$ and $h(B,A) = \max_{b \in B} \min_{a \in A} \|b-a\|$, respectively. $A$ and $B$ are a pair of point sets. HD is automatic and less labor-intensive. Note that this study utilized a 100% HD metric, unlike earlier work (Archip, 2007), which featured a 95% HD metric.

In (Drakopoulos et al., 2021), an extensive study for registration accuracy was performed using 30 cases and anatomical landmarks selected by a neurosurgeon, as suggested in Hastreiter et al., 2004. The neurosurgeon located six landmarks in each registered preoperative and corresponding intraoperative image volume. Landmarks A and B were selected individually in the cortex near the tumor; C and D were selected at the anterior horn and the triangular part of the lateral ventricle, respectively; E and F were selected at the junction between the pons and mid-brain and the roof of the fourth ventricle, respectively (see Figure 11).

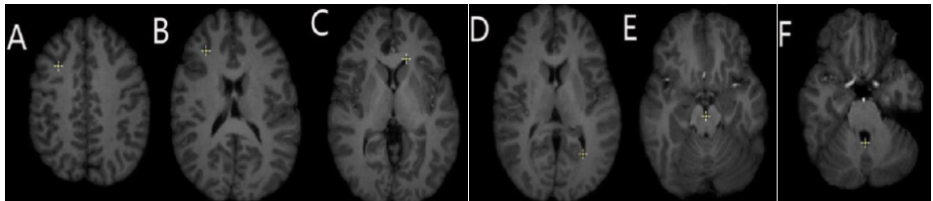

Figure 11. Locations of the anatomical landmarks (A - F) used for the quantitative evaluation of the registration accuracy by a neurosurgeon: A, B: cortex near tumor, C: anterior horn of later ventricle, D: triangular part of later ventricle, E: junction between pons and mid-brain, F: roof of the fourth ventricle.

The neurosurgeon located one and four additional landmarks of functional interest individually. For each case, these additional landmarks were selected depending on the location of the tumor, the surgical approach, and the visibility of the preoperative and intraoperative images. These structures of functional interest include, amongst others, the primary motor cortex, the pyramidal tract, the Sylvian fissure, the lateral border of the thalamus, the basal ganglia, the posterior limb of the internal capsule, and significant vessels. The error was calculated for each landmark as the distance between the landmark location in the registered preoperative image and the corresponding intraoperative image. Table 2 compares NEM and ANRR with the rest of the publicly available registration methods: RR, B-Spline, and PBNRR. It presents each case's minimum, maximum, and mean errors. The assessment confirms that the NEM and ANRR provide the most accurate registration, with an average minimum error of 1.36 mm and 1.03 mm, an average mean error of 3.69 mm and 3.22 mm, and an average max error is 7.79 and 6.59 mm, respectively.

Table 2. Quantitative registration results using six anatomical landmarks (A-F). The average minimum, maximum, and mean errors are computed over twenty-five cases for NEM and five additional (i.e., thirty cases) for rigid registration (RR), PBNRR, and ANRR.

| Method | Average min error (mm) | Average max error (mm) | Average mean error (mm) |
|---|---|---|---|
| RR | 3.19 | 8.90 | 5.60 |
| BSPLINE | 2.15 | 8.29 | 4.40 |
| PBNRR | 1.11 | 6.81 | 3.47 |
| NEMNRR | 1.36 | 7.79 | 3.69 |
| ANRR | 1.03 | 6.59 | 3.22 |

*3.3 Real-time Adaptive Image-To-Mesh Conversion*

Although numerous grid or mesh generation methods have been presented in the literature to date, only some can deal in real-time with NRR requirements. Some methods generate low-fidelity meshes with a very large number of elements without allowing control to the user as far as the mesh size. In contrast, an I2M conversion method initially developed for NRR (Fedorov et al., 2005) and improved by (Liu et al., 2010) by meshing multi-tissue segmented images and in (Drakopoulos et al., 2015) by addressing mesh gradation (i.e., control mesh size without compromising the fidelity of the mesh). Figure 12 shows the use of a finite element mesh for non-rigid registration of brain parenchyma and tumor tissues using tetrahedral elements. The resulting meshes are not smooth since, for NRR, the objective is to capture the details of a segmented image, where the fidelity and control of mesh size are critical, as opposed to blood flow applications where smoothness is important (Kazakidi et al., 2016).

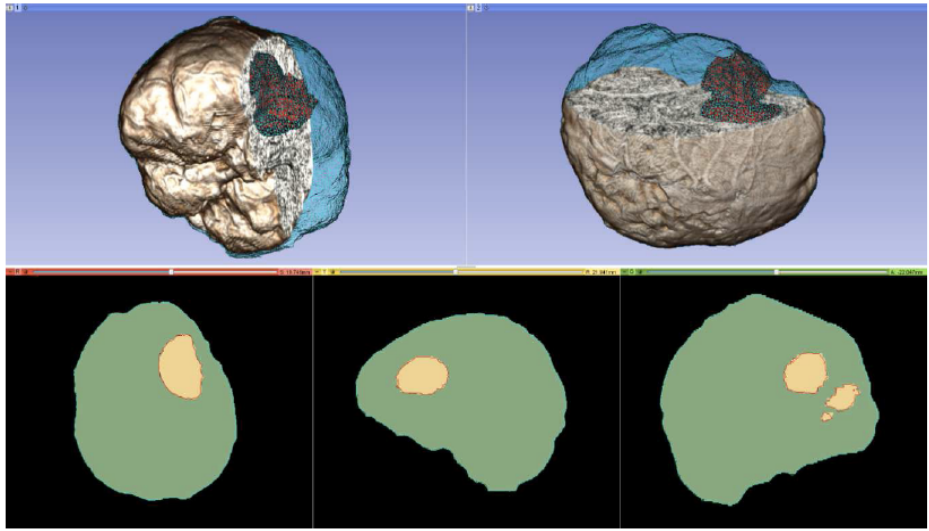

Figure 12. Top row: the mesh superimposed on a volume rendering of the MRI data. Cyan and red represent the brain parenchyma and tumor meshes, respectively. Bottom-row: mesh fidelity illustrated on an axial, sagittal, and coronal slice.

A detailed analysis of requirements for I2M conversion appears in (Fedorov and Chrisochoides, 2008); next, we list a subset:

1. *Suboptimal distribution of the registration points*: Concerning mesh elements: a small number of points makes the numerical formulation more sensitive to outliers and introduces additional displacement error due to integral voxel displacements recovered by block matching. The distribution of points also influences the condition of the $[K+H^TSH]$ matrix.
2. Fidelity to Image Boundaries: The mesh must provide a reasonably close representation of the underlying object in the image. A mesh, which provides 100% fidelity, would be the one whose boundaries exactly fit the piecewise-linear boundary between the voxels of the image corresponding to the tissue and all other voxels. However, in many cases, such a mesh would provide an extremely fine level of detail and contain an excessively large number of elements.
3. *Gradation of the Mesh*: Interpolation error can be reduced by having smaller size mesh elements; at the same time, this leads to a larger number of elements and a longer solution time. Thus, it is essential to selectively adapt (refine) the mesh in the areas of interest.
4. *Size of the Mesh*: The application imposes tight time constraints on the solution process, and the solution time depends on the size of the linear system. It is crucial to bind the number of degrees of freedom (i.e., vertices of the mesh) and keep it to the minimum necessary to achieve the desired solution accuracy.
5. *The shape of Mesh Elements*: Very large or small angles can increase the condition number of the stiffness matrix and lead to slower convergence rates for the iterative solver. The condition number can be improved with preconditioning; one could use a diagonal precondition (fast but less effective), and thus, the shape of elements is still very critical.
6. *Mesh Generation (or Refinement) Time*: The set of registration points is dynamically changing through the execution of the algorithm as outlier points are discarded. This can lead to a few registration points assigned to a mesh vertex, violating the requirement (1). To circumvent this problem, the mesh must be refined/coarsened, or a new mesh must be generated (re-mesh) as necessary.
7. *Multi-tissue Capabilities:* *This work conjectures that as more tissues are incorporated into the model, such as the falx of the brain, the higher the registration accuracy will be more research and evaluation will be required to validate this hypothesis experimentally.*

A detailed review of I2M conversion methods and evaluation of their performance w.r.t FEM-based non-rigid registration method appears in (Foteinos et al., 2010). This study indicates that BCC-based methods (Liu et al., 2012) are preferable to Delaunay-based methods (Foteinos et al., 2014) and Lattice Decimation methods (Chernikov and Chrisochoides, 2011) because the relatively dense initial BCC mesh captures the object surface without much compression, thus preserving the good angles of the BCC triangulation. However, the main critique of BCC-based methods is still on the (large) number of elements required for a given fidelity. This issue will be addressed by introducing mesh gradation while addressing the optimal distribution constraint.

*3.3.1 Distribution of Registration Points*

Six of the seven requirements for NRR are generic to many other applications, and their treatment is outside the scope of this chapter. Instead, the rest of this section will focus on the (sub-)optimal distribution of registration points. As noted earlier, the NRR software pipeline utilizes BCC-based and real-time Delaunay-based image-to-mesh conversion for mesh generation. The Delaunay-based approach is by far more flexible in terms of mesh gradation. It can generate a mesh that faithfully captures (with geometric guarantees) the surface of the input image and the interface between the two tissues. However, it does not consider any information about the registration points recovered by the block-matching step.

In previous work (Fedorov, 2008), the distribution of landmarks over the mesh was incorporated into the mesh generation module using custom sizing functions for two different mesh generation methods (Delaunay refinement and Advancing Front). These modifications aim to equidistribute the landmarks among the mesh elements; this is expected to improve the registration error. The evaluation presented in (Fedorov, 2008) was based on synthetic deformation fields and showed that these modifications, to a degree, reduce the registration error.

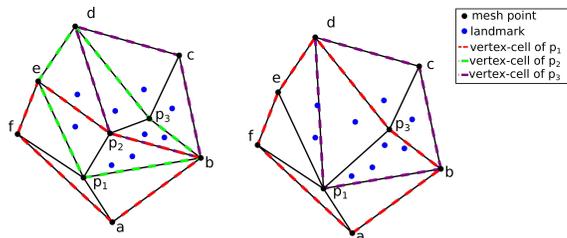

Figure 13 Optimizing landmark distribution.

In a subsequent effort (Tsolakis, 2021) with preliminary results appearing in (Drakopoulos et al., 2021), the same sizing function is applied to validate the method's effectiveness on a retrospective case study. Preliminary results from applying mesh adaptation methods that originate from the Computational Fluid Dynamics field are presented in this paper. For completeness, a summary of the method employed in (Fedorov, 2008) is presented along with the modifications that can turn it into an anisotropic metric-based method. The (sub-)optimal distribution of the registration points can be formulated as assigning approximately the same number of registration points at each mesh vertex cell complex, where a mesh vertex cell complex is defined as the set of all the elements attached to a vertex. See, for example, Figure 13. The p1, p2, and p3 vertex cells on the left have 3, 7, and 5 landmarks, respectively. While on the right, by collapsing edge $p_2p_1$, one attempts to equidistribute the landmarks. Both the vertex cells of $p_1$ and $p_2$ *now have* seven landmarks.

The crux of the method is to set the local spacing at each vertex equal to the distance to the k-th closest registration point. Assuming an ideal spacing, each vertex's mesh vertex cell complex will contain k registration points. An illustration for k = 5 is given in Figure 14 left. Notice that another way to interpret the sizing constraint at each vertex is using a sphere centered at each mesh vertex with a radius equal to the distance to the k-th registration point.

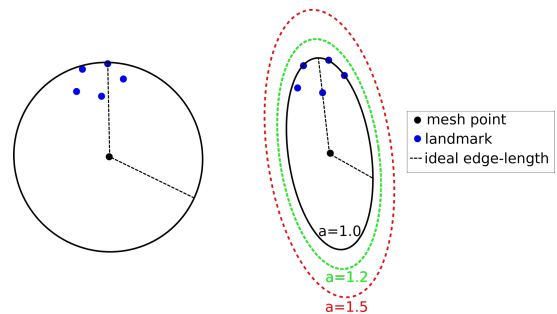

Figure 14 Left: Isotropic metric that sets the spacing equal to the distance of the fifth closest registration point. Right: Anisotropic metric based on the five registration points for different values of the inflation constant.

This technique produces adaptive meshes but does not efficiently capture the local distribution of landmarks. This is because only the k-th point is used, and the relative positions of the other k-1 landmarks are disregarded. To improve this, one can substitute the spheres at each vertex with the smallest bounding ellipsoid that encompasses the k closest registration points and is centered at the vertex. Describing the local spacing as an ellipsoid gives the ability to capture the local distribution of the landmarks better due to the increased degrees of freedom of an ellipsoid compared to a sphere (see, for example, Figure 14 right).

Creating the minimum volume ellipsoid that encloses a given pointset is a problem well studied in the convex optimization literate. The constructed ellipsoid has a natural mapping to a 3x3 positive definite matrix that can be used as a metric that guides the anisotropic mesh adaptation procedure. An additional flexibility to the mesh adaptation procedure can be introduced by an 'inflation' (constant $a$), which is introduced and is common for all the points; it allows to enlarge all ellipsoids by a constant factor. The goal of this parameter is to allow the mesh generation procedure to perform operations that may not conform to the strict size but improve the overall result. See Figure 14, right.

To incorporate the above approach to ANRR, the mesh generated by the Parallel Optimistic Delaunay Mesh (Foteinos et al., 2014) at each iteration, along with the landmarks identified by the Block-Matching step, are used to build a metric field. The metric field is constructed by iterating in parallel the mesh vertices and evaluating the k-closest registration points using a k-nn search from the VTK library (*VTK - The Visualization Toolkit*, 2020).

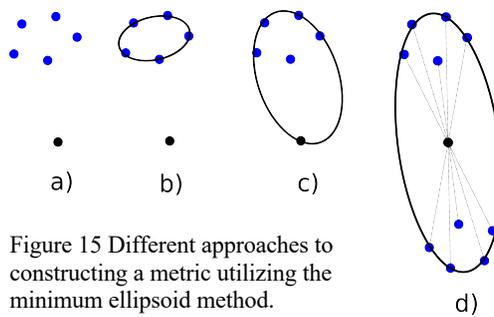

Figure 15 Different approaches to constructing a metric utilizing the minimum ellipsoid method.

The minimum volume bounding ellipsoid is constructed using the Khachiyan algorithm. Directly using the landmarks (Figure 15 (b)) will not yield an ellipse centered at a mesh point. Including the mesh point into the input of the minimum ellipsoid algorithm does not fix the issue (see Figure 15 (c)). Instead, one can generate reflections of the k-closest landmarks by the mesh point and include them in the input of the minimum ellipsoid algorithm. Due to symmetry, the mesh point will always be in the center of the constructed ellipsoid. Finally, the mesh is adapted using MMG3D (Dapogny et al., 2014) using the metric field derived from the constructed ellipsoids. Figure 14 depicts the difference of isotropic vs. anisotropic and sup-optimal mesh for a single case. Notice that the number of elements generated constrains the anisotropy; it must be approximately equal to the number of elements in the isotropic mesh.

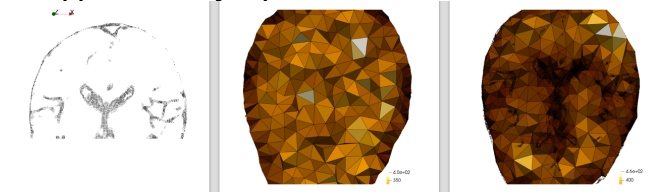

Figure 15. Registration points (left), isotropic mesh (center) and anisotropic sub-optimal mesh (right).

Table 3 presents preliminary data from two new cases: (A: case 9) from (Drakopoulos et al., 21), provided by HSH (male, with a glioma at Left Frontal location of the brain, where Partial Resection is performed, with preop-MRI and iMRI image sizes and spacing: 448x512x176 and 0.488x0.488x1.00, respectively) and (B: case 18) from (Drakopoulos et al., 21), provided by HSH (female, with glioma at Left Frontal location of the brain, where Total Resection is performed, with preop-MRI and iMRI image sizes and spacing: 448x512x176 and 0.488x0.488x1.00, respectively).

Table 3. Hausdorff Distance (HD) and error using A-F landmarks reported in mm. Where baseline uses the default I2M within ANRR, isotropic uses the equidistribution method by (Fedorov, 2008), and anisotropic with different values for the alpha weight (in parenthesis).

| Case A | HD | Min error | Max error | Mean error | # tets | # vertices |
|---|---|---|---|---|---|---|
| baseline | 2.24 | 1.07 | 5.90 | 3.51 | 13,210 | 3,264 |
| isotropic | 1.95 | 1.22 | 7.53 | 3.71 | 19,893 | 4,177 |
| anisotropic (a=1.0) | 2.22 | 0.55 | 7.85 | 3.99 | 22,383 | 4.520 |
| anisotropic (a=1.2) | 2.00 | 1.01 | 7.10 | 3.70 | 17.593 | 3.629 |
| anisotropic (a=1.5) | 2.64 | 0.93 | 6.15 | 3.25 | 13,291 | 2,838 |

| Case B | HD | Min error | Max error | Mean error | # tets | # vertices |
|---|---|---|---|---|---|---|
| baseline | 4.06 | 2.06 | 5.37 | 3.65 | 11,040 | 2,833 |
| isotropic | 3.42 | 2.29 | 5.76 | 3.92 | 19,946 | 4,008 |
| anisotropic (a=1.0) | 3.71 | 2.12 | 5.50 | 3.96 | 22,342 | 4,460 |
| anisotropic (a=1.2) | 4.05 | 2.06 | 5.05 | 3.61 | 18,077 | 3.766 |
| anisotropic (a=1.5) | 4.05 | 1.92 | 5.17 | 3.65 | 13,812 | 2,983 |

In contrast, for case A, the HD error for Rigid Registration (RR), B-Spline, and PBNRR (without optimal distribution of registration points) is 10.59mm, 5.28mm, and 10.76mm, respectively. While for case B the HD error for RR, B-Spline, and PBNRR (without optimal distribution of registration points) is: 25.72mm, 25.72mm, and 23.90mm, respectively. In both cases, the optimal distribution within the ANRR method reduced the error to about five to six times compared to RR and PBNRR. While the error using A-F landmarks is improved (see Table 2), more work is needed.

## 4. Discussion

This paper addressed some of the computational aspects of a D$^4$NRR for Image Guided Neurosurgery. This D$^4$NRR method was used for the first time ever in clinical practice (Chrisochoides, 2006) and (Archip et al., 2007); it completed and made available non-rigid registration results in the OR (at BWH) during tumor resection procedures using image landmark tracking across the entire brain volume. Example cases used for the evaluation are publicly available (NCIGT, 2021). In addition, many of the registration codes used (RR, BSplines, PBNRR) are publicly available within Slicer3D.

Before any conclusions are drawn, clarification is important. The automatic method (Garlapati, 2013) was used to evaluate the registration accuracy quantitatively because it is fast and does not require manual intervention. It relies on Canny edge detection (Canny, 1986) to compute two-point sets. The first point set is computed from the preoperative volume and then transformed (using the deformation field computed by each registration method) from the preoperative to the intraoperative space. The second point set is computed from the intraoperative volume. A Hausdorff Distance (HD) metric (Commandeur, 2011) was employed to calculate the degree of displacement between the two-point sets. This approach is crude and helpful only for a relative comparison of the methods and should not be used in any clinical setting. However, the NEMNRR method, the way is formulated, can provide (in the future) a way to compute the correspondence between those two sets of points for the HD error, making HD error much more reliable.

All registration methods (RR, B-Splines, PBNRR, NEMNRR, ANRR, and its recent version for the sub-optimal distribution of registration points) can be fine-tuned and optimized even more. There is space for improvement with the use of Cloud computing and Deep Learning, as suggested in (Drakopoulos et al., 2021). Tables 4 and 5 (Appendix I) depict the most important parameters the authors considered in this study. As it is apparent and experienced by this group, for a single case to search — even a reduced parameter space from Tables 4 and 5 — using 400 cores (about 10 HPC nodes), running a single adaptive physics-based registration method takes a full month. The computational cost is prohibitive! This suggests that using Deep Learning for patient-specific parameter values will be beneficial; some preliminary results appeared in (Drakopoulos et al., 2021). In summary, Deep Learning improved the accuracy of ANRR by finding optimal values for its parameters. On average, APBNRR with deep learning is ~8.45 times better than rigid registration, ~6.71 times better than B-Spline registration, and ~7.9 times better than PBNRR, an older version of ANRR. Again, more work is required to improve the accuracy of cases with deep brain tumors and validate the current state of the software.

A final and perhaps more important outcome of this work was workforce development. Six Ph.D. students trained and delivered these results: (1) Dr. Fedorov initiated this research with his work on "Enabling Technology for Non-Rigid Registration during Image-Guided Neurosurgery" (Fedorov, 2009), currently an Associate Professor at the Department of Radiology, Brigham and Women's Hospital, Harvard Medical School; (2) Dr. Liu improved by introducing the NEM and the multi-tissue heterogeneous model with his work "On the Real-Time Performance, Robustness, and Accuracy of Medical Image Non-Rigid Registration" (Liu, 2011), currently Senior Researcher at Tencent; (3) Dr. Drakopoulos then improved even further by following a pure geometric approach to manage tumor resection and substantially improved many topologic aspects of the multi-tissue mesh generation method with his work on "On the Geometric Modeling for Aerospace and Health Care Applications" (Drakopoulos, 2017), currently a R&D Staff Engineer at Synopsis Inc. However, the real-time NRR in the case of ANRR only becomes feasible because of the work by (4) Dr. Foteinos on "Real-Time Image to Mesh Conversion for Finite Element Simulations" (Foteinos, 2013), currently a Meshing specialist at Altair and (5) Dr. Tsolakis' work on "Unified Framework for Parallel Anisotropic Mesh Adaptation" (Tsolakis, 2021) is promising in addressing the optimal distribution of registration points and potentially help improve the accuracy of ANRR; Dr. Tsolakis currently is Senior R&D Engineer. However, none of this would be possible without the tenacity and persistence of (6) Dr. Kot to untangle the original FEM NRR code and his early contributions to many software systems aspects (Kot, 2011), Dr. Kot is a Software Development Engineer at Intel Corporation.

In conclusion, this multi-disciplinary and multi-national team at CRTC, BWH, and INRIA demonstrated more than 15 years ago that DDDAS deformable registration can be used in real-time for image-guided neurosurgery. The accuracy of the specific deformable registration is improving to a point within reach to achieve the 1mm goal for the registration error. Deep learning, for using patient-specific parameters in NEMNRR and ANRR, can help to get closer to this goal. However, some issues remain unresolved (Angelopoulos and Chrisochoides, 2018): "First, more training data needs to be collected to allow the deep learning model to offer more accurate predictions. Second, work needs to be done to enable the deep learning model to generate a parameter pool that is limited in size and can also be evaluated rapidly. Finally, the accuracy of ANRR needs to be further improved, especially the performance of ANRR regarding deep tumors which involve very large brain deformation."

## 5. Future Work

The CRTC team will focus on using Quantum Computing to address some of the combinatorial aspects of the NRR problem. CRTC's recent work on Quantum Hadamard Edge Detection (QHED), one of the simplest but one of the most computationally intensive image processing kernels, suggests that a hybrid (classical and quantum) model might be promising (Billias and Chrisochoides, 2023), but still far away from clinical use.

### 5.1 Preliminary Results on Quantum Edge Detection

A promising quantum image processing algorithm, Quantum Hadamard Edge Detection (X-W. Yao et al., 2017). However, resultant circuit depth becomes exponential for the image encoding portion of the algorithm. Due to noise in NISQ-era quantum computers and exponential memory requirements, large medical images (such as pre-op and intraoperative brain images, see Figure 1) must be decomposed into sub-images and processed in parallel and individually to address fidelity issues. However, the image decomposition scheme proposed in (X-W. Yao et al., 2017) can cause false edges to appear across the output image. In addition, the use of decrement permutation in NISQ-era circuits requires a very large number of multi-controlled NOT (MCX) gates, which results in an exponential number of controlled not (CX) gates for the mapping of the QHED circuit onto the quantum computer hardware. In short, both the image encoding, and the edge detection parts of the algorithm produce exponential circuit depth, which compounds a massive loss in fidelity. To ensure correct boundary detection, we use classical space-filling curves commonly used in parallel numeric computations (Chrisochoides et al., 1997) to correct the artificial edges during the pre and postprocessing of input vectors. Figure 17 depicts our approach, for more details, see (Billias et al., 2023).

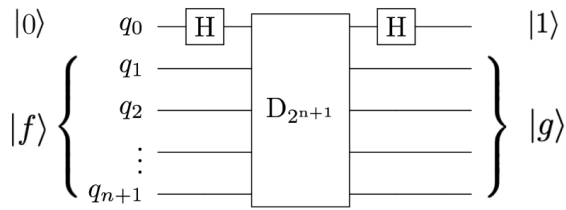

Figure 17. The QHED circuit proposed in (Yao et al., 2017) with an ancillary qubit. The $D_2^{n+1}$ gate is a type of amplitude permutation that acts as a decrement operation on the input state vector.

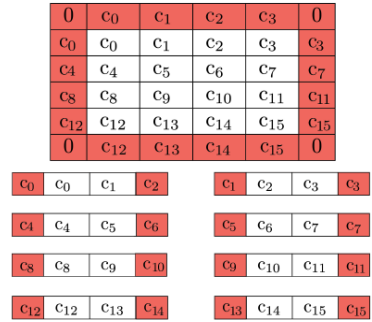

Figure 16. Buffer pixels of mirrored value are applied to the boundary of the overall image (0's for the corners, as they won't be used). Further decomposition is possible by adding neighboring pixels to be used as a buffer due to QHED error. The red cells are ultimately disposed of in the final output.

To reduce the number of CX gates, we use a linear number of ancillary qubits. We also address fidelity concerns with optimization techniques from (J.K Ferris et al., 2022) to minimize hardware noise in both the amplitude encoding and QHED circuits.

Optimizing the topology and software of quantum circuits can improve results on physical hardware. A simulated noisy backend from IBM (A. Abbas et al., 2020) is utilized to evaluate the results of the proposed QHED optimizations (see

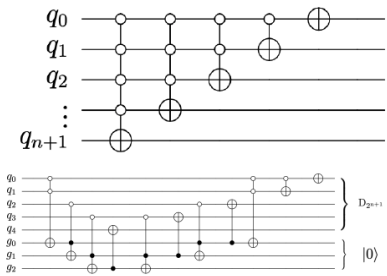

Figure 18. The circuit proposed in (X-W. Yao et al., 2017) for decrementing using n+1 qubits requires a descending series of MCX gates (top), which must be converted into an exponential series of CNOT gates to perform the decrement permutation on real hardware. An alternate decrement circuit (bottom) utilizes only CX and Toffoli gates, transpiling into a linear number of total CX gates.

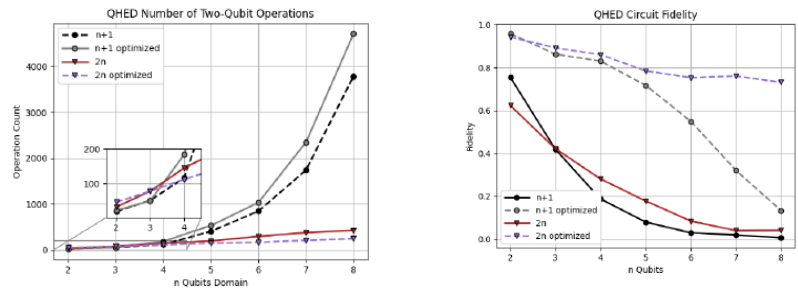

Figure 19. Analysis of the QHED algorithm using IBM's backend for both the proposed (red and blue) and (Yao et al, 2017) circuits (black and gray lines). Each of the two circuits is analyzed with and without full optimizations.

Figure 18). An additional list of all optimizations is presented in (Billias et al., 2023). Based on our analysis of Figure 19, even with the optimizations we made to the original circuit, we still need a near-exponential amount of CX operations with respect to an increasing number n of data encoding qubits due to the amplitude encoding portion of the circuit. This results in a rapid loss of fidelity for any n > 5 number of data encoding qubits. Although quantum technology is not ready today for clinical use, ongoing efforts in quantum hardware and software, according to quantum computing experts put this technology on our radar "screen". According to Quantum Technology Monitor (April 2023) "four industries likely to see the earliest [practical and] economic impact from quantum computing—automotive, chemicals, financial services, and life sciences—stand to potentially gain up to $1.3 trillion in value by 2035" (McKinsey & Company, 2023). For the near future we will focus on understanding the challenges in quantum medical image computing and Digital Health broadly and investing in workforce development.

In summary, we improved the Quantum Edge Detection method (Yao et al., 2017) to generate comprehensible results on NISQ hardware for Medical Imaging use cases. Our priority is addressing the pre-and post-processing and the quantum edge detection circuit itself. To do so, we implement buffer pixels to eliminate image artifacts, a new decrement permutation circuit, optimizations for realistic images on today's QPUs, and additional optimization techniques to improve circuit fidelity and reduce the depth and the number of two-qubit operations.

**Acknowledgments:**  AF, YL, PF, AK, FD, CT, and NC were supported in part by the NSF grants: CCS-0719929, CCS0750901, CCF-0833081, CCF-1439079, NASA grant no. NNX15AU39A and Modeling and Simulation Fellowship at Old Dominion University. In addition, the Richard T. Cheng Endowment partially supports NC and EB, and NC is supported by the John Simon Guggenheim Foundation.  AF was supported in part by a grant from the Brain Science Foundation. RK, AG, and PB were supported in part by NIH grants: U41 RR019703, P01 CA67165, NIH R01NS049251, NIH 5R01EB027134, and NCIGT P41EB015898. This research was supported by an allocation through the TeraGrid Advanced Support Program. This work was performed [in part] using computational facilities at Old Dominion University and the College of William and Mary (in Virginia) enabled by grants from the National Science Foundation and Virginia's Commonwealth Technology Research Fund. The content is solely the authors' responsibility and does not necessarily represent the government agencies' official views supporting.

# Appendix I

Table 4. Parameters used in this study for rigid registration (RR) and B-Spline non-rigid registration methods implemented in 3D Slicer. MMI: Mattes Mutual Information; VR3DT: Versor Rigid 3D Transform; LBFGSB: Limited memory Broyden Fletcher Goldfarb Shannon minimization with Simple Bounds.

| Parameter | Value | | Description |
| --- | --- | --- | --- |
| | **RR** | **B-Spline** | |
| Cost Metric | MMI | MMI | Mattes Mutual Information |
| Interpolation Mode | Linear | Linear | |
| Sampling percentage | 5% | 5% | Percentage of image voxels sampled for MMI |
| Histogram bins | 100 | 100 | Number of histogram levels |
| Optimizer type | VR3DT | LBFGSB | - |
| Max number of iterations | 1500 | 1500 | Maximum number of iterations for optimizer |
| Grid Size | - | 15x15x15 | Number of subdivisions of the B-Spline Grid |
| Min step length | $10^{-3}$ | $10^{-3}$ | Min threshold step for optimizer |
| Projected Gradient Tolerance | - | $10^{-5}$ | Used by LBFGSB |

Table 5. A few important parameters are used for PBNRR, NEMNRR, and ANRR.

| Parameter | Value | Description |
| --- | --- | --- |
| Initialization transform | Rigid | Rigid transformation to initialize the non-rigid registration |
| Connectivity pattern | "face" | Pattern for the selection of blocks |
| $F_s$ | 5% | % selected blocks from total number of blocks |
| $B_{s,x} \times B_{s,y} \times B_{s,z}$ | $3 \times 3 \times 3$ | Block size (in voxels) |
| $W_{s,x} \times W_{s,y} \times W_{s,z}$ | $7 \times 7 \times 3$ (BS) $9 \times 9 \times 3$ (PR) $13 \times 13 \times 3$ (TR, STR) | Block matching window size (in voxels) |
| $\delta$ | 5 | Mesh size (PBNRR, ANRR) |
| $E_b$ | 2.1 KPA | Young's modulus for brain parenchyma |
| $E_t$ | 21 KPA | Young's modulus for tumor |
| $v_b$ | 0.45 | Poisson ratio for brain tumor and parenchyma |
| $v_t$ | 0.1 | Poisson ratio for ventricle (NEMNRR) |
| $F_r$ | 25% | % of rejected outlier blocks |
| $N_{rej}$ | 10 | Number of outlier rejection steps |
| $N_{iter,max}$ | 10 | Max number of adaptive iterations (ANRR) |

E. Billias and N. Chrisochoides on the scalability of Quantum Edge Detection for NISQ-era Imaging Simulations. CRTC Technical Report, 2023.

Quantum Technology Monitor, (accessed August 2023), https://www.mckinsey.com/capabilities/mckinsey-digital/our-insights/quantum-technology-sees-record-investments-progress-on-talent-gap#/